\documentclass[conference]{IEEEtran}
\IEEEoverridecommandlockouts
\usepackage{cite}
\usepackage{amsmath,amssymb,amsfonts}
\usepackage{graphicx}
\usepackage{textcomp}
\usepackage{xcolor}
\usepackage{listings}
\usepackage{moresize}
\usepackage{pdfpages}
\usepackage{attachfile2}
\usepackage{hyperref}
\usepackage{amsmath, amsthm} % Load math packages
\usepackage{algorithm}
\usepackage{algpseudocode}
\usepackage{microtype}
\usepackage{booktabs,tabularx}
\usepackage{multirow}
\usepackage{xcolor}
\usepackage[T1]{fontenc}
\usepackage{listings}
\usepackage{xcolor}
\usepackage{listings}

\lstdefinelanguage{Dafny}{
  morekeywords={datatype,predicate,method,returns,ensures,requires,var,while,invariant,if,forall,exists,in,seq,nat,int},
  sensitive=true,
  morecomment=[l]{//},
}
\lstdefinestyle{err}{language={},basicstyle=\ttfamily\scriptsize\color{black!70},keywordstyle={}}
\lstdefinestyle{diff}{language={},basicstyle=\ttfamily\scriptsize,keywordstyle={},
  moredelim=[il][\color{red!70!black}]{--},moredelim=[il][\color{green!45!black}]{++}}

\usepackage[most]{tcolorbox}
\newtcolorbox{rqanswer}[1]{enhanced, colback=gray!8, colframe=gray!60,
  boxrule=0.5pt, arc=1pt, left=4pt, right=4pt, top=3pt, bottom=3pt,
  before upper={\noindent\textbf{RQ#1 Answer:}\ }}
\newcommand{\rqhead}[2]{\par\smallskip\noindent\textbf{RQ#1:} \textit{#2}\par\smallskip}

\theoremstyle{definition} 
\newtheorem{definition}{Definition}

\usepackage{soul}
\usepackage{tikz}
\usetikzlibrary{calc}

\makeatletter
\newif\if@anonymize

\@anonymizefalse  % Uncomment to show text

\if@anonymize
  \newcommand{\highlight@DoHighlight}{
    \fill [outer sep = -15pt, inner sep = 0pt, color=white]
          ($(begin highlight)+(0,8pt)$) rectangle ($(end highlight)+(0,-3pt)$) ;
  }

  \newcommand{\highlight@BeginHighlight}{
    \coordinate (begin highlight) at (0,0) ;
  }

  \newcommand{\highlight@EndHighlight}{
    \coordinate (end highlight) at (0,0) ;
  }

  \newdimen\highlight@previous
  \newdimen\highlight@current
  \newlength{\item@width}

  \DeclareRobustCommand*\anonymize{%
    \SOUL@setup
    \def\SOUL@preamble{%
      \begin{tikzpicture}[overlay, remember picture]
        \highlight@BeginHighlight
        \highlight@EndHighlight
      \end{tikzpicture}%
    }%
    \def\SOUL@postamble{%
      \begin{tikzpicture}[overlay, remember picture]
        \highlight@EndHighlight
        \highlight@DoHighlight
      \end{tikzpicture}%
    }%
    \def\SOUL@everyhyphen{%
      \discretionary{%
        \SOUL@setkern\SOUL@hyphkern
        \SOUL@sethyphenchar
        \tikz[overlay, remember picture] \highlight@EndHighlight ;%
      }{%
      }{%
        \SOUL@setkern\SOUL@charkern
      }%
    }%
    \def\SOUL@everyexhyphen##1{%
      \SOUL@setkern\SOUL@hyphkern
      \settowidth{\item@width}{##1}%
      \makebox[\item@width]{}%
      \discretionary{%
        \tikz[overlay, remember picture] \highlight@EndHighlight ;%
      }{%
      }{%
        \SOUL@setkern\SOUL@charkern
      }%
    }%
    \def\SOUL@everysyllable{%
      \begin{tikzpicture}[overlay, remember picture]
        \path let \p0 = (begin highlight), \p1 = (0,0) in \pgfextra
          \global\highlight@previous=\y0
          \global\highlight@current =\y1
        \endpgfextra (0,0) ;
        \ifdim\highlight@current < \highlight@previous
          \highlight@DoHighlight
          \highlight@BeginHighlight
        \fi
      \end{tikzpicture}%
      \settowidth{\item@width}{\the\SOUL@syllable}%
      \makebox[\item@width]{}%
      \tikz[overlay, remember picture] \highlight@EndHighlight ;%
    }%
    \SOUL@
  }
\else
  \newcommand{\anonymize}[1]{#1}
\fi
\makeatother

\usepackage{xcolor,pifont}
\newcommand*\colourcheck[1]{%
  \expandafter\newcommand\csname #1check\endcsname{\textcolor{#1}{\ding{52}}}%
}
\newcommand*\colourxmark[1]{%
  \expandafter\newcommand\csname #1xmark\endcsname{\textcolor{#1}{\ding{56}}}%
}
\colourcheck{green}
\colourxmark{red}

\def\BibTeX{{\rm B\kern-.05em{\sc i\kern-.025em b}\kern-.08em
    T\kern-.1667em\lower.7ex\hbox{E}\kern-.125emX}}
\begin{document}

\title{Learning from Mutants: Teaching Small Language Models to Synthesise Loop Invariants\\

\thanks{\anonymize{This work was conducted with the financial support of the Science Foundation Ireland Centre for Research Training in Digitally-Enhanced Reality (d-real) under Grant No. 18/CRT/6224 and the ERC Consolidator Grant ARTIST 101002685. For the purpose of Open Access, the authors  applied a CC BY public copyright licence to any Author Accepted Manuscript version arising from this submission.}}
}

\author{

\IEEEauthorblockN{\anonymize{Daragh King}}
\IEEEauthorblockA{\anonymize{\textit{School of Computer Science and Statistics}}} 
\textit{\anonymize{Trinity College Dublin}}\\
\anonymize{Dublin, Ireland} \\
\anonymize{{kingd6@tcd.ie}}

\and
\IEEEauthorblockN{\anonymize{Vasileios Koutavas}}
\IEEEauthorblockA{\textit{\anonymize{School of Computer Science and Statistics}} \\
\textit{\anonymize{Trinity College Dublin}}\\
\anonymize{Dublin, Ireland} \\
\anonymize{Vasileios.Koutavas@tcd.ie}}

}

\maketitle

\begin{abstract}
Large language models (LLMs) have proven promising for synthesising loop invariants, a long-standing obstacle to automated program verification. Yet, frontier LLMs are typically accessible only through third-party APIs, hampering reproducibility and local deployment. Small language models (SLMs) sidestep these obstacles but require task-specific fine-tuning, for which high-quality training data is needed. We present a failure-driven data curation method for fine-tuning language models to synthesise loop invariants. Inspired by mutation testing, our approach applies semantic mutation operators to ground-truth invariants, producing \emph{semantic mutants} that induce verification failure. Each mutant is paired with verifier feedback and its ground-truth counterpart, yielding training examples for feedback-guided invariant refinement. Whereas the vast space of mutants is a long-standing burden in mutation testing, it becomes an asset when generating training data, multiplying the examples obtainable from a fixed set of verified programs. We implement our approach for Dafny, augmenting an existing synthetic dataset of verified programs, and introduce a framework that iteratively refines candidate invariants using verifier feedback. On 472 DafnyBench programs, our best-performing fine-tuned Gemma-4-E4B configuration verifies 50.8\% of programs zero-shot, compared to 8.1\% for the base model, a 6.3$\times$ improvement. Higher-order mutants prove more effective as training data than first-order mutants, and adding mutation-based examples to direct synthesis examples, in which invariants are removed rather than mutated, improves performance under iterative refinement. Our approach also transfers to Qwen3.5-4B, whose zero-shot verification rate rises from 9.3\% to 53.6\%.
\end{abstract}


\section{Introduction}\label{sec:intro}
Loop invariants are inductive assertions used to establish the partial correctness of looping computations in Hoare-style program logics~\cite{Floyd67,hoare-logic}. Loop invariant synthesis is a central challenge in automated program verification and is undecidable in general. Static approaches~\cite{cousot1977abstract,humenberger2017invariant} to this challenge typically perform well in their targeted problem domains, but can fail to generalise. Dynamic approaches~\cite{ernst2007daikon,hangal2002tracking,nguyen2012using} infer likely invariants from program executions instead of reasoning about program semantics, but their effectiveness is constrained by the quantity and quality of these execution traces. Data-driven approaches~\cite{sharma2013data,garg2014ice,padhi2016data,nguyen2017counterexample} address this by checking learned candidates with a verifier and refining them on its counterexamples.

More recently, the use of large language models (LLMs) for loop invariant synthesis and related specification tasks has emerged as a promising alternative~\cite{pei2023can,janssen2024can,pascoal2025automatic,king2025llm,king2026llms}, but utilising such models presents its own set of challenges. Frontier LLMs are typically proprietary and accessible only through third-party APIs, so using them means sending code and data to the companies that host them~\cite{wang2025comprehensive, hanke2024open}. Additionally, reproducibility suffers from rapid progress in proprietary LLM development. For example, early works applying OpenAI's GPT-4 to invariant synthesis~\cite{kamath2023finding} can no longer be reproduced, as the model was withdrawn from OpenAI's API in 2026~\cite{openai_deprecations}. Open-weight LLMs address some of these concerns, but their substantial compute requirements often preclude local deployment~\cite{wang2025comprehensive, belcak2025small}.

Small language models (SLMs) avoid these issues. Belcak et al.~\cite{belcak2025small} define SLMs as models that can perform inference with practical latency on common consumer hardware, and consider most models with fewer than 10 billion parameters to be SLMs. Because SLMs can run on the user's own hardware, code never leaves the user's machine and no provider can withdraw access. Off the shelf, however, they perform poorly at invariant synthesis; in our evaluation, base Gemma-4-E4B and Qwen3.5-4B verify fewer than 10\% of test programs zero-shot, and even with thinking enabled Gemma-4-E4B verifies only 26.7\%. They therefore need task-specific fine-tuning~\cite{wang2025comprehensive}, for which high-quality datasets are essential~\cite{pinto2026not}. Existing datasets of verified programs~\cite{yan2025re,poesia2026formal} provide correct invariants, but not the failing invariants and verifier feedback needed to train a model to repair its own candidate proposals. DafnyPro~\cite{banerjee2026dafnypro} and ATLAS~\cite{baksys2025atlas} do train on failed attempts with error feedback, but they obtain these failures by running language models.

In this work, we target loop invariant synthesis for Dafny~\cite{leino2010dafny}, an auto-active verifier in which the programmer annotates code with specifications and proof hints, and an SMT solver discharges the resulting proof obligations. Programs with loops usually need loop invariants among these hints. Automating the synthesis of invariants is therefore a valuable objective, recently pursued with LLMs (e.g.\ \cite{pascoal2025automatic,breen2026axdafny}). Such approaches inherit the limitations of LLMs outlined above. Fine-tuned open models of 7 to 14 billion parameters~\cite{poesia2024dafny,banerjee2026dafnypro,baksys2025atlas} avoid some of these limitations, and our work builds on this direction with a new source of training data.

We present a failure-driven data curation method for fine-tuning language models to synthesise invariants in Dafny. Our approach relies upon failure-inducing \emph{semantic mutants} of ground-truth invariants, i.e.\ mutations of invariants which induce verification failure. These mutant invariants are paired with verifier feedback from Dafny and mapped to their ground-truth counterparts, yielding training examples for feedback-guided invariant refinement.

Our approach is inspired by mutation testing~\cite{jia2010analysis}. We define 26 semantic mutation operators which, like those used in mutation testing, can be applied repeatedly to produce higher-order mutants~\cite{jia2009higher}. In mutation testing, and in mutation-based invariant inference~\cite{galeotti2015inferring}, the resulting blow-up in the number of mutants is a fundamental obstacle. When generating training data, this abundance instead becomes an asset: each verification-failing mutant yields a training example, multiplying the data obtainable from a fixed set of verified programs. We complement these mutation-based examples with \emph{direct synthesis} examples, in which an invariant is removed rather than mutated, and train models for use within an invariant refinement framework that iteratively repairs candidate invariants using verifier feedback.

We evaluate our approach by fine-tuning Gemma-4-E4B~\cite{team2026gemma} on datasets derived from Formal Disco~\cite{poesia2026formal}, and testing on 472 programs from DafnyBench~\cite{loughridge2024dafnybench}. Fine-tuning substantially improves performance: our best-performing Gemma configuration verifies 50.8\% of test programs zero-shot, compared to 8.1\% for the base model, a 6.3$\times$ improvement. Higher-order mutants prove more effective than first-order mutants as training data, and adding mutation-based examples to direct synthesis examples improves performance under iterative refinement. The approach also transfers to a second model family: fine-tuning Qwen3.5-4B~\cite{qwen35} on the same data raises its zero-shot verification rate from 9.3\% to 53.6\%.

In summary, our contributions are:
\begin{itemize}
    \item A failure-driven data curation method that uses semantic mutation operators to generate verification-failing invariants for fine-tuning (Section~\ref{data curation}).
    \item An invariant refinement framework for Dafny that iteratively repairs candidate invariants using verifier feedback (Section~\ref{sec:refinement}).
    \item An evaluation on DafnyBench across two model families, examining the effects of training data composition and LoRA rank (Section~\ref{sec:results}).
\end{itemize}

The remainder of this paper is organised as follows. Section~\ref{sec:background} provides background on Hoare logic, mutation testing, and SLMs. Section~\ref{method} describes our invariant refinement framework and data curation method. Section~\ref{sec:exp setup} details our experimental setup, Section~\ref{sec:results} presents our results, and Section~\ref{sec:threats} discusses threats to validity. Section~\ref{sec:related} discusses related work, and Section~\ref{sec:conclusion} concludes.

\section{Background}\label{sec:background}
\subsection{Hoare Logic, Loop Invariants, and Dafny}
Hoare logic~\cite{hoare-logic} provides a formal system for reasoning about the correctness of imperative programs. Fundamental to this logic is the Hoare triple (Definition~\ref{HoareTriple}), which relies on pre- and postconditions: postconditions specify the desired property of the program state upon termination, whilst preconditions specify the required property of the initial program state for it to behave as intended (i.e.\ to satisfy the program's postcondition). Hoare logic has total and partial correctness versions, where total also considers termination as part of the intended behaviour and partial does not.\footnote{Dafny proves total correctness by default, using loop variants. We take loop variants as given and focus on invariant synthesis and partial correctness.}

\begin{definition}[Hoare Triple, Partial Correctness]\label{HoareTriple}
A Hoare triple $\{P\}\ C\ \{Q\}$ asserts that if the precondition $P$ holds before
executing command $C$, and $C$ terminates, then the postcondition $Q$ holds after
execution.
\end{definition}

Verifying programs containing loops requires auxiliary \emph{loop invariant} predicates.
Broadly speaking, loop invariants are assertions that hold before and after each loop iteration. In the context of Hoare logic, a loop invariant sufficient for verification must satisfy the following three conditions:  

\begin{definition}[Initiation]\label{initiation-def} The invariant $I$ must hold on loop entry. Equivalently, if $P$ holds immediately before the loop, then: \begin{equation} P \Rightarrow I \end{equation} \end{definition}

\begin{definition}[Preservation]\label{preservation-def}
If the invariant $I$ holds and the loop condition $B$ is true, executing the loop body $C$
must preserve the invariant. Equivalently:
\begin{equation}
\{I \land B\}\ C\ \{I\}
\end{equation}
\end{definition}

\begin{definition}[Exit]\label{exit-def}
When the invariant $I$ holds and the loop exits (i.e.\ $B$ is false), the property $Q$ required after the loop
must be satisfied.
\begin{equation}
I \land \neg B \Rightarrow Q
\end{equation}
\end{definition}

Dafny~\cite{leino2010dafny} is a verification-aware programming language that supports the automatic verification of formally specified programs. Building on Hoare-style reasoning, Dafny translates programs into verification conditions that are discharged automatically by the Z3 SMT solver~\cite{de2008z3}. Loops, however, must be manually annotated with loop invariants: Dafny checks that each invariant holds on loop entry (Definition~\ref{initiation-def}) and is preserved by the loop body (Definition~\ref{preservation-def}), and then relies on the invariants, together with the negated loop condition, to verify the code that follows the loop (Definition~\ref{exit-def}). An invariant that fails any of these checks causes verification to fail.

\subsection{Mutation Testing}\label{Mutation Testing}
Mutation testing is a fault-based testing technique used to assess the quality of a program's test suite~\cite{jia2010analysis}. Syntactically altered versions of the program, known as \emph{mutants}, are generated by applying \emph{mutation operators}: rules that make small syntactic changes, or \emph{mutations}, e.g.\ replacing one arithmetic operator with another. The test suite is then run against each mutant; if some test distinguishes the mutant from the original program, the mutant is \emph{killed}, otherwise it is said to \emph{survive}. Surviving mutants may indicate weaknesses in a program's test suite, though some may instead be \emph{equivalent mutants}, i.e.\ mutants semantically equivalent to the original program and therefore undetectable by any test. Determining whether a mutant is equivalent is undecidable in general, making equivalent mutants a long-standing obstacle in mutation testing~\cite{jia2010analysis}.

Higher-order mutation testing extends this approach by applying multiple mutations to the original program to create each mutant~\cite{jia2009higher}. A mutant produced by $q$ mutations is termed a $q$th-order mutant, with conventional mutants being first-order. It has been argued that higher-order mutants better simulate real software faults~\cite{harman2010manifesto}. However, the space of possible mutants grows combinatorially with order, making exhaustive higher-order mutation testing computationally prohibitive~\cite{jia2009higher}.

\subsection{Small Language Models and Fine-Tuning}
As noted in Section~\ref{sec:intro}, we follow Belcak et al.~\cite{belcak2025small} in considering SLMs to be language models that can perform inference with practical latency on common consumer hardware; as of writing, these typically have fewer than 10 billion parameters. Due to their smaller size and compute requirements, SLMs lend themselves well to task-specific fine-tuning and local deployment~\cite{wang2025comprehensive}. SLMs can also approach the performance of LLMs on specialised software engineering tasks; on requirements classification, for example, they come close to the F1 scores of LLMs~\cite{zadenoori2025does}.

Language models are often fine-tuned using low-rank adaptation (LoRA)~\cite{hu2021lora}, or its memory-efficient variant, quantised LoRA (QLoRA)~\cite{dettmers2023qlora}. LoRA freezes the pretrained weights of the model and instead trains small low-rank matrices injected into selected layers. This substantially reduces the number of trainable parameters, and can obviate the need for full-parameter fine-tuning. QLoRA further reduces memory requirements by quantising the frozen base model to 4-bit precision while training the LoRA matrices on top.

Using LoRA-based methods to fine-tune SLMs places task-specific adaptation within reach of modest hardware. For instance, QLoRA enables fine-tuning of a 65-billion-parameter model on a single 48GB GPU~\cite{dettmers2023qlora}, and models an order of magnitude smaller can comfortably be fine-tuned on a single consumer-grade GPU. The resulting adapters are also compact, typically a small fraction of the base model's size~\cite{hu2021lora}, allowing multiple task-specific adapters to be stored, shared, and swapped over a single base model. Together, these properties extend the accessibility and local deployability that motivate the use of SLMs. However, effective fine-tuning still requires high-quality training data, whose generation is non-trivial~\cite{pinto2026not}.

\section{Method}\label{method}

\begin{table*}[t]
\centering
\caption{Mutation operators, grouped by category.}
\label{tab:operators}
\scriptsize
\setlength{\tabcolsep}{4pt}
\begin{tabular}{@{}l l l@{}}
\toprule
Category & Operator & Edit \\
\midrule
\multirow{9}{*}{Deletion}
 & \texttt{drop\_clause}             & Drop one non-bound clause. \\
 & \texttt{drop\_progress\_clause}   & Drop a clause about a variable the loop body assigns. \\
 & \texttt{drop\_equality\_clause}   & Drop a top-level equality. \\
 & \texttt{drop\_init\_fact}         & Drop an equality that restates a pre-loop assignment. \\
 & \texttt{drop\_several\_clauses}   & Drop two or more non-bound clauses, keeping at least one. \\
 & \texttt{weaken\_to\_bounds\_only} & Keep only the bound clauses. \\
 & \texttt{drop\_bound\_clause}      & Drop one bound clause. \\
 & \texttt{drop\_inner\_conjunct}    & Drop one conjunct from a conjunction within a clause. \\
 & \texttt{drop\_chained\_side}      & Drop one side of a chained bound, such as \texttt{a <= x <= b}. \\
\midrule
\multirow{8}{*}{Bounds}
 & \texttt{loosen\_bound}              & Relax a comparison by one step (e.g.\ \texttt{<} becomes \texttt{<=}). \\
 & \texttt{tighten\_upper\_bound}      & Tighten an upper bound by one step (e.g.\ \texttt{<=} becomes \texttt{<}). \\
 & \texttt{strict\_lower\_bound}       & Make an inclusive lower bound strict. \\
 & \texttt{raise\_lower\_bound}        & Raise an inclusive lower bound by one. \\
 & \texttt{close\_open\_bound}         & Turn a one-sided bound into a two-sided range. \\
 & \texttt{add\_accumulator\_bound}    & Add a bound on an integer accumulator the loop body assigns. \\
 & \texttt{replace\_bound\_expression} & Replace a bound expression with another in-scope expression. \\
 & \texttt{guard\_as\_invariant}       & Add the \texttt{while} guard as an invariant clause. \\
\midrule
\multirow{4}{*}{Quantifiers}
 & \texttt{strengthen\_quantifier}     & Make a strict quantifier bound inclusive (\texttt{<} becomes \texttt{<=}). \\
 & \texttt{widen\_quantifier\_range}   & Widen a quantifier range. \\
 & \texttt{narrow\_quantifier\_range}  & Narrow a quantifier range. \\
 & \texttt{mutate\_quantifier\_body}   & Change the body of a quantifier. \\
\midrule
\multirow{3}{*}{Borrowed clauses}
 & \texttt{initializer\_as\_invariant}    & Add a pre-loop assignment \texttt{v == E} as an invariant clause. \\
 & \texttt{borrow\_sibling\_loop\_clause} & Copy a ground-truth clause from a sibling loop. \\
 & \texttt{relation\_mutation}            & Rewrite a top-level relation, keeping its left-hand side. \\
\midrule
\multirow{2}{*}{Ill-formed syntax}
 & \texttt{unbind\_identifier}    & Introduce a name that is not in scope at the loop head. \\
 & \texttt{surface\_syntax\_slip} & Write surface syntax that Dafny rejects. \\
\bottomrule
\end{tabular}
\end{table*}

This section describes our approach in two parts: an invariant refinement framework, in which a language model iteratively repairs candidate invariants using verifier feedback (Section~\ref{sec:refinement}), and a mutation-based data curation method, which produces training examples matching the framework's expected input and output formats (Section~\ref{data curation}).

\subsection{Invariant Refinement Framework}\label{sec:refinement}
We define \emph{invariant refinement} as the iterative process of improving a candidate loop invariant until it is sufficient for verification.

Consider a program $\mathcal{P}$ consisting of a set of methods $M$, in which every loop is initially annotated with an empty invariant, i.e.\ one equivalent to \texttt{true}. Using a program verifier and a language model, our invariant refinement framework traverses $\mathcal{P}$ method by method, carrying out the following steps for each $m_j \in M$, up to a maximum of $k$ refinement attempts per method:
\begin{enumerate}
    \item Attempt to verify $m_j$. If verification succeeds, proceed to $m_{j+1}$. Otherwise, proceed to step 2.
    \item Inspect $m_j$ for loops. If $m_j$ contains no loops, proceed to $m_{j+1}$, as $m_j$'s verification failure is out of scope for invariant refinement. Otherwise, proceed to step 3.
    \item Construct the invariant refinement prompt $\textsc{Prompt}(\mathcal{P}, m_j, \hat{I}_j, e)$ from the program $\mathcal{P}$, the method $m_j$, its current candidate invariants $\hat{I}_j$ (initially empty), and the associated verification error $e$.
    \item Prompt the language model with $\textsc{Prompt}(\mathcal{P}, m_j, \hat{I}_j, e)$ and parse its response into new candidate invariants, which replace $\hat{I}_j$.
    \item Create $m'_j$ by replacing the loop invariants of $m_j$ with $\hat{I}_j$, and return to step 1 with $m'_j$.
\end{enumerate}
If the $k$ attempts are exhausted without $m_j$ (or its patched variants) verifying, the framework proceeds to $m_{j+1}$. The program is verified once every method in $M$ verifies.

We implement the above for Dafny, whose modular verification makes method-by-method traversal sound: each method is verified against the specifications of the methods it calls, rather than their bodies, so refining the invariants of one method does not affect the verification of another.

The refinement prompt (an example is included in our supplementary material~\cite{supp}) consists of fixed instructions followed by step-specific inputs. The instructions specify the response format, a single JSON object listing proposed invariant expressions for each loop, together with syntactic constraints (e.g.\ no statements or edits to method specifications) and guidance for common verification failures, such as weakening a clause that fails on loop entry. The step-specific inputs comprise the loops of $m_j$ that may be edited, the current candidate invariants $\hat{I}_j$ in the same JSON format, the verifier output $e$, and the program as sent to the verifier, with $\hat{I}_j$ in place and line numbers matching the verifier's diagnostics. When there are no previous candidates, as on the first attempt, $\hat{I}_j$ is given as an empty proposal (\texttt{\{"proposals": []\}}). Each proposed expression is inserted as an \texttt{invariant} clause on its loop, and the proposals for a method replace its previous candidates rather than amending them. Loops outside $m_j$ are locked: proposals for them are ignored, and their current invariants are restored automatically. Each response is thus applied to the invariant-stripped original program, with the invariants of locked loops restored, rather than to the previous candidate program.

\begin{figure*}[t]
\centering
\definecolor{codebg}{gray}{0.96}
\definecolor{coderule}{gray}{0.75}
\definecolor{hl}{RGB}{0,90,160}
\definecolor{del}{RGB}{170,30,30}
\definecolor{ins}{RGB}{20,120,40}
\lstset{language={}, basicstyle=\ttfamily\scriptsize, columns=fullflexible, keepspaces=true,
  backgroundcolor=\color{codebg}, frame=single, rulecolor=\color{coderule},
  framesep=4pt, framerule=0.4pt, xleftmargin=4.4pt, xrightmargin=4.4pt,
  aboveskip=3pt, belowskip=3pt,
  moredelim={[is][\color{hl}]{@}{@}},
  moredelim={[is][\color{del}]{`}{`}},
  moredelim={[is][\color{ins}]{~}{~}}}
\newcommand{\panel}[1]{\par\noindent{\footnotesize\textbf{#1}}\par}
\begin{minipage}[t]{0.485\textwidth}
\panel{(a) Ground-truth program}
\begin{lstlisting}
method ApplyView(items: seq<Item>, config: ViewConfig)
  returns (result: seq<Item>)
  ensures forall i :: 0 <= i < |result| ==>
            ItemMatchesView(result[i], config)
  ensures forall i :: 0 <= i < |result| ==>
            result[i] in items
{
  result := [];
  var idx := 0;
  while idx < |items|
    @invariant 0 <= idx <= |items|@
    invariant forall j :: 0 <= j < |result| ==>
                ItemMatchesView(result[j], config)
    @invariant forall j :: 0 <= j < |result| ==>@
    @            result[j] in items@
  {
    if ItemMatchesView(items[idx], config) {
      result := result + [items[idx]];
    }
    idx := idx + 1;
  }
}
\end{lstlisting}
\end{minipage}\hfill
\begin{minipage}[t]{0.485\textwidth}
\panel{(b) First-order mutant: \texttt{tighten\_upper\_bound}}
\begin{lstlisting}
`- invariant 0 <= idx <= |items|`
~+ invariant 0 <= idx < |items|~
\end{lstlisting}
\begin{lstlisting}[basicstyle=\ttfamily\scriptsize\color{black!65}]
Error: this invariant could not be proved to be
       maintained by the loop
Error: this loop invariant could not be proved
       on entry
\end{lstlisting}
\vspace{4pt}
\panel{(c) Second-order mutant: \texttt{surface\_syntax\_slip} on (b)}
\begin{lstlisting}
`- invariant forall j :: 0 <= j < |result| ==>`
`-             result[j] in items`
~+ invariant all (0 <= j < |result| ==>~
~+             result[j] in items)~
\end{lstlisting}
\begin{lstlisting}[basicstyle=\ttfamily\scriptsize\color{black!65}]
Error: unresolved identifier: all
Error: unresolved identifier: j   (reported twice)
\end{lstlisting}
\end{minipage}
\caption{Mutation example. (a)~A training program with its ground-truth loop invariants (definitions of \texttt{Item}, \texttt{ViewConfig}, and \texttt{ItemMatchesView} omitted); the clauses mutated in (b) and (c) are highlighted. (b)~A first-order mutant produced by \texttt{tighten\_upper\_bound}, which fails verification because the tightened bound does not hold on loop entry when \texttt{items} is empty, nor after the final iteration. (c)~A second-order mutant, produced by applying \texttt{surface\_syntax\_slip} to (b), which fails to resolve. Verifier errors are abridged. In both cases, the training target is the ground-truth invariant in (a).}
\label{fig:mutation}
\end{figure*}

\subsection{Data Curation}\label{data curation}

Inspired by mutation testing (Section~\ref{Mutation Testing}), we devise a data curation method for fine-tuning language models to synthesise loop invariants. Whereas mutation testing alters program code to produce mutant programs, our approach instead alters a program's proof annotations, specifically loop invariants. To do so, we apply \emph{semantic mutation operators} to alter the logical meaning of invariants, producing \emph{semantic mutants} (see Figure~\ref{fig:mutation} for an example). In this work, we define 26 mutation operators (Table~\ref{tab:operators}), grouped into five categories: deletion, bounds, quantifiers, borrowed clauses, and ill-formed syntax.\footnote{Strictly speaking, ill-formed syntax operators do not alter an invariant's logical meaning, but instead render it unparsable or ill-scoped. We include them because language models can produce such errors, making these mutants useful training examples.}

Consider a verified program $\mathcal{P}$ and one of its ground-truth invariants $I$, comprising all invariant clauses annotating one loop. Let $O$ denote the set of semantic mutation operators. Mutants retained for $I$ are stored in a set $R_I$ of tuples $(I', \mathcal{P}', e)$, where $I'$ is the mutant invariant, $\mathcal{P}'$ is the program with $I$ replaced by $I'$, and $e$ is the error produced when verifying $\mathcal{P}'$. To produce up to $N_m$ first-order semantic mutants of $I$, we carry out the following steps, up to a maximum of $K_m$ generation attempts:
\begin{enumerate}
    \item Choose an operator $o \in O$ uniformly at random from those not yet chosen for $I$. Apply $o$ to $I$ to produce a candidate mutant $I'$.
    \item Create $\mathcal{P}'$ by replacing $I$ with $I'$ in $\mathcal{P}$, leaving the program's remaining (ground-truth) invariants unchanged.
    \item Verify $\mathcal{P}'$, obtaining verification error $e$ on failure. If $\mathcal{P}'$ verifies, or verification times out, discard $I'$ and $\mathcal{P}'$ and return to step 1. Otherwise, proceed to step 4; since $\mathcal{P}$ verifies, the verification failure of $\mathcal{P}'$ is attributable to the semantic mutation.
    \item If $I'$ is syntactically identical to the mutant of any tuple in $R_I$, discard $I'$ and $\mathcal{P}'$ and return to step 1. Otherwise, add $(I', \mathcal{P}', e)$ to $R_I$.
    \item If $|R_I| = N_m$, stop. Otherwise, return to step 1.
\end{enumerate}
Every attempt counts towards the budget $K_m$, including those in which the chosen operator is inapplicable to $I$; generation therefore stops once $N_m$ mutants have been retained or the $K_m$ attempts are exhausted. As operators are chosen without replacement, $K_m \leq |O|$.

Higher-order mutants are produced in separate, successive phases. Each phase applies the aforementioned process to mutants of the previous order: each $(q-1)$th-order mutant serves as the input in place of its ground-truth invariant, producing $q$th-order mutants, while all other invariants in the program again retain their ground-truth form. Consequently, even though the input program no longer verifies, any verification failure remains attributable to the mutated invariant. Training targets, however, are always the original ground-truth invariants, regardless of mutant order. Figure~\ref{fig:mutation} shows an example of first- and second-order semantic mutants and the verification errors they induce.

Retained tuples of every order are used to create training examples for language model fine-tuning. For each $(I', \mathcal{P}', e) \in R_I$, with $m$ the method of $\mathcal{P}'$ containing $I'$, we construct the refinement prompt $\textsc{Prompt}(\mathcal{P}', m, I', e)$, in which the mutant $I'$ takes the place of the model's previous candidate invariants. This prompt is then paired with the original ground-truth invariant, expressed in the response format the framework parses. As verification is modular, $m$ is the only method of $\mathcal{P}'$ that fails, so $e$ matches the error the refinement framework would observe. Training examples therefore match both the prompts the model receives and the responses it is expected to produce during refinement.

\section{Experimental Setup}\label{sec:exp setup}
This section describes the setup used to evaluate our approach, which addresses the following research questions:

\rqhead{1}{How is Gemma's ability to synthesise correct loop invariants affected by fine-tuning on differing orders of mutation-based training data, and how does this compare to fine-tuning using direct synthesis examples?}

\rqhead{2}{How does scaling training data by combining mutants of different orders affect Gemma's ability to synthesise correct loop invariants?}

\rqhead{3}{How does combining direct synthesis and mutation-based training data affect Gemma's ability to synthesise correct loop invariants, compared with either data source alone?}

\rqhead{4}{How does increasing the LoRA rank and alpha affect Gemma's ability to synthesise correct loop invariants when fine-tuned on combined mutation-based and direct synthesis training data?}

\rqhead{5}{How does fine-tuning on combined mutation-based and direct synthesis training data affect the ability of a different model family, Qwen, to synthesise correct loop invariants, and how does this compare to Gemma?}

We first describe the construction of the training datasets and the fine-tuning procedure, followed by the models and configurations evaluated. We then describe the test dataset, along with the prompting, refinement framework, and verification settings used during evaluation. Section~\ref{sec:results} addresses each research question in turn.

\subsection{Training Datasets}

\subsubsection{Source}
The source training dataset $\mathcal{S}$ consists of $1600$ verified programs randomly sampled from Formal Disco~\cite{poesia2026formal}, where each program contains at least one loop, and each loop has ground-truth invariants. $\mathcal{S}$ is used as a basis to create further datasets for model fine-tuning.

\subsubsection{Mutation-Based}
Mutation-based datasets are created from $\mathcal{S}$ using the failure-driven data curation method described in Section~\ref{data curation}, with $K_m = 10$ and mutant orders $q \in \{1, 2, 4\}$.\footnote{Third-order mutants are generated as an intermediate step towards fourth-order mutants, but are not used for fine-tuning.} First-order mutants are generated with $N_m = 4$, yielding approximately 16k examples, which are randomly subsampled to approximately 6k. Each higher-order phase then uses $N_m = 1$, producing at most one mutant per example of the previous order, so that all mutation-based datasets are of comparable size. We denote the dataset of $q$th-order mutants by $M_q$.

\subsubsection{Direct Synthesis}
The direct synthesis dataset $D$ is created from $\mathcal{S}$ in a similar manner to the mutation-based datasets (Section~\ref{data curation}), except that each ground-truth invariant is removed rather than mutated. As removal yields a single candidate per invariant, each ground-truth invariant in $\mathcal{S}$ produces at most one training example. For methods containing several loops, the remaining loops retain their ground-truth invariants in these examples, whereas at test time every loop starts from an empty invariant.

\subsection{Training Hyperparameters}
Unless explicitly stated otherwise, models are fine-tuned for $1$ epoch using QLoRA~\cite{dettmers2023qlora}, with 4-bit quantisation of the base model, a LoRA rank and alpha of 8, and a dropout of 0. We use the AdamW optimiser~\cite{loshchilov2017decoupled} with a learning rate of $2 \times 10^{-4}$, 5 warmup steps, a linear learning-rate schedule, and an effective batch size of 2. The same settings are used for each model trained, with the exception of one Gemma configuration trained with a LoRA rank and alpha of 32 (Section~\ref{model variants}).

\subsection{Model and Dataset Variants}\label{model variants}
Our experiments primarily use Gemma-4-E4B~\cite{team2026gemma} (hereafter \emph{Gemma}), a dense model with approximately 4.5 billion effective and 8 billion total parameters. To assess whether our approach transfers across model families, we also evaluate Qwen3.5-4B~\cite{qwen35} (hereafter \emph{Qwen}), a model with approximately 4 billion parameters. As our training datasets are model-agnostic, the same datasets are used to fine-tune both models.

Table~\ref{tab:configs} lists the configurations evaluated. $G'$ and $Q'$ denote the base Gemma and Qwen models respectively, and $G'_T$ denotes base Gemma with ``thinking'' enabled. All other configurations denote a model fine-tuned on the dataset(s) indicated in the subscript, with combined datasets marked by $+$; a numeric superscript denotes a LoRA rank and alpha differing from the default of 8.

\begin{table}[t]
\centering
\caption{Model configurations evaluated.}
\label{tab:configs}
\footnotesize
\setlength{\tabcolsep}{3pt}
\begin{tabular}{@{}l l c r@{}}
\toprule
Configuration & Training data & Rank$^{a}$ & Size$^{b}$ \\
\midrule
\multicolumn{4}{@{}l}{\textbf{Gemma-4-E4B}} \\
\addlinespace[2pt]
$G'$                     & None                            & --- & --- \\
$G'_T$                   & None (thinking)                 & --- & --- \\
\addlinespace[3pt]
$G_{D}$                  & Direct synthesis                & 8   & 3k \\
$G_{M_1}$                & 1st-order mutants               & 8   & 6k \\
$G_{M_2}$                & 2nd-order mutants               & 8   & 6k \\
$G_{M_4}$                & 4th-order mutants               & 8   & 6k \\
$G_{M_1+M_2+M_4}$        & 1st-, 2nd-, 4th-order mutants   & 8   & 18k \\
$G_{D+M_1+M_2+M_4}$      & Direct synthesis + all mutants  & 8   & 21k \\
$G^{32}_{D+M_1+M_2+M_4}$ & Direct synthesis + all mutants  & 32  & 21k \\
\midrule
\multicolumn{4}{@{}l}{\textbf{Qwen3.5-4B}} \\
\addlinespace[2pt]
$Q'$                     & None                            & --- & --- \\
\addlinespace[3pt]
$Q_{D+M_1+M_2+M_4}$      & Direct synthesis + all mutants  & 8   & 21k \\
\bottomrule
\multicolumn{4}{@{}l}{\scriptsize $^{a}$LoRA rank and alpha.} \\
\multicolumn{4}{@{}l}{\scriptsize $^{b}$Number of training examples, rounded to the nearest thousand.} \\
\end{tabular}
\end{table}

\subsection{Test Dataset}\label{test dataset}
The test dataset comprises the $472$ programs from DafnyBench~\cite{loughridge2024dafnybench} that contain at least one loop and whose loop invariants are necessary for verification under the settings defined in Section~\ref{verification settings}, i.e.\ test programs do not verify without their loop invariants. During evaluation, all loop invariants are removed from all test programs, and models must synthesise them from empty invariants. All other annotations, including loop variants (\texttt{decreases} clauses), are retained, so that termination does not depend on synthesised annotations. This setting follows Faria et al.~\cite{pascoal2025automatic}; the original DafnyBench task instead removes assertions as well as loop invariants~\cite{loughridge2024dafnybench}.

\subsection{Prompting}
To aid reproducibility, we use greedy decoding (temperature~$0$) when prompting all models, removing sampling variance between experiments. Thinking is disabled for all configurations except $G'_T$, whose thinking is not capped by a token budget; $G'_T$ therefore generates substantially more tokens per prompt than the other configurations.

\subsection{Refinement Framework Parameters}\label{framework parameters}
Each configuration is evaluated in two settings. In the \emph{zero-shot} setting, each eligible method\footnote{Eligible methods contain at least one loop and fail verification once invariants are removed; methods that verify without any invariants are excluded, as they require no invariant synthesis.} receives a single attempt ($k = 1$): the model is prompted once, and the resulting candidate invariants are verified. For both base and fine-tuned models, \emph{zero-shot} thus denotes a single attempt, without the opportunity to refine the model's own candidates using verifier feedback. In the \emph{iterative} setting, the invariant refinement framework (Section~\ref{sec:refinement}) is applied with a budget of $k = 5$ refinement attempts per method, including the initial attempt.

\subsection{Verification Settings}\label{verification settings}
All verification is performed using Dafny 4.11.0 with its bundled Z3 SMT solver (version 4.12.1).\footnote{\raggedright Each call is \texttt{dafny verify -{}-verification-time-limit 30 -{}-cores 100\% -{}-isolate-assertions -{}-allow-warnings}.} Each assertion is verified in its own batch, subject to a 30-second time limit, and any batch exceeding it is treated as a failure. Each Dafny invocation is also subject to a 60-second wall-clock timeout, and invocations exceeding it are likewise treated as failures. The same settings are used both for data curation and for evaluation.

\section{Results}\label{sec:results}
\begin{figure*}[t]
\centering
\includegraphics[width=\textwidth]{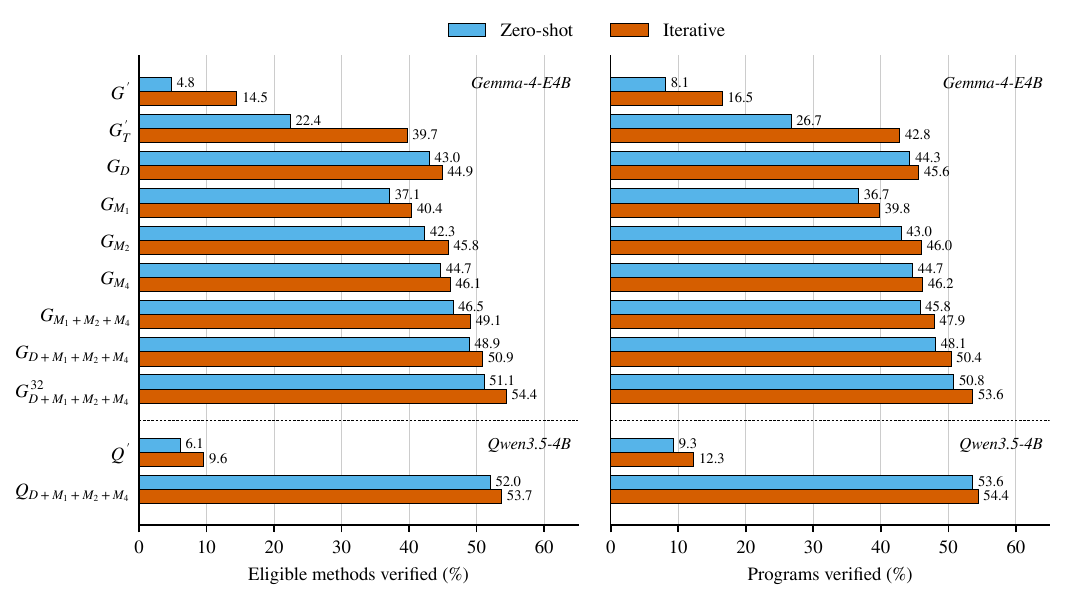}
\caption{Percentage of eligible methods (left) and programs (right) verified on DafnyBench by each configuration, under zero-shot generation and iterative invariant refinement. Primes denote base models; all other configurations are fine-tuned on the datasets indicated in their subscripts (Table~\ref{tab:configs}).}
\label{fig:results}
\end{figure*}

The following presents the results obtained using the experimental setup described in Section~\ref{sec:exp setup}, addressing each of the research questions posed there in turn. Figure~\ref{fig:results} reports the percentage of eligible methods (Section~\ref{framework parameters}) and programs verified by each configuration. As the two metrics largely follow the same trends, we discuss program-level results unless stated otherwise, reporting zero-shot and iterative results as $x$/$y$. We test differences between configurations using exact McNemar tests on per-program outcomes, applying Holm correction across the nine pairwise comparisons reported below, separately for each setting, and describe a difference as significant if its corrected $p < 0.05$; all $p$-values reported below are Holm-corrected.

\noindent\textbf{RQ1:} Fine-tuning on any single training dataset substantially improves Gemma's performance. The base model $G'$ verifies 8.1\%/16.5\% of programs, whereas every configuration fine-tuned on a single dataset verifies over four times as many programs zero-shot. Even fine-tuning on first-order mutants alone ($G_{M_1}$) improves significantly on the base model in both settings ($p < 0.001$).

Mutant order has a clear effect: with each dataset containing approximately 6k examples derived from the same first-order mutants, performance increases from 36.7\%/39.8\% for $G_{M_1}$, to 43.0\%/46.0\% for $G_{M_2}$, and 44.7\%/46.2\% for $G_{M_4}$, an ordering that also holds at the method level. Most of this gain occurs between first and second order, with improvements of 6.3/6.2 percentage points, which are significant in both settings, compared to 1.7/0.2 between second and fourth order, which are not. This echoes the argument from mutation testing that higher-order mutants better resemble real faults~\cite{harman2010manifesto}, although simpler explanations are also possible: higher-order mutants lie further from their targets, so each example requires reproducing more of the ground truth, and their mix of operators and resulting errors differs from that of first-order mutants.

Fine-tuning on direct synthesis examples is similarly effective: $G_D$ verifies 44.3\%/45.6\% of programs, comparable to $G_{M_2}$ and $G_{M_4}$, with no significant difference from $G_{M_4}$, despite being trained on roughly half as many examples. Notably, every single-dataset configuration outperforms Gemma with thinking enabled ($G'_T$, 26.7\%) in the zero-shot setting, and all but $G_{M_1}$ also outperform it under iterative refinement (42.8\%), without incurring any reasoning overhead.

\begin{rqanswer}{1}
Fine-tuning on mutation-based data improves Gemma's zero-shot program verification rate by up to 5.5$\times$, from 8.1\% ($G'$) to 44.7\% ($G_{M_4}$). Second-order mutants significantly outperform first-order mutants, while the further gain from fourth-order mutants is not significant. Second- and fourth-order mutants ($G_{M_2}$, 43.0\%/46.0\%; $G_{M_4}$, 44.7\%/46.2\%) perform comparably to direct synthesis examples ($G_D$, 44.3\%/45.6\%), though the latter achieve this with half as many examples.
\end{rqanswer}

\noindent\textbf{RQ2:} Combining first-, second-, and fourth-order mutants ($G_{M_1+M_2+M_4}$, $\sim$18k examples) yields 45.8\%/47.9\%, improving on the best single-order configuration, $G_{M_4}$, by 1.1/1.7 percentage points (1.8/3.0 at the method level). The gain is consistent across both metrics (programs and methods verified) and both settings (zero-shot and iterative refinement), but modest relative to the threefold increase in training data, and not statistically significant. As combining datasets increases both the volume and the diversity of examples, these results do not isolate the contribution of either factor.

\begin{rqanswer}{2}
Combining mutants of different orders ($G_{M_1+M_2+M_4}$, 45.8\%/47.9\%) yields small gains over the best single-order configuration ($G_{M_4}$, 44.7\%/46.2\%) across both metrics and settings, but these are not statistically significant, despite the threefold increase in training data.
\end{rqanswer}

\noindent\textbf{RQ3:} Supplementing the combined mutation data with direct synthesis examples ($G_{D+M_1+M_2+M_4}$) raises performance to 48.1\%/50.4\%, an improvement of 2.3/2.5 percentage points over $G_{M_1+M_2+M_4}$ (2.4/1.8 at the method level), which is not significant, and of 3.8/4.8 over $G_D$ alone, which is significant under iterative refinement ($p = 0.010$; $p = 0.080$ zero-shot). As $G_D$ and $G_{D+M_1+M_2+M_4}$ are trained on data derived from the same source programs, adding mutation-based examples improves the performance obtainable from a fixed set of verified programs, although this comparison does not control for the number of training steps (Section~\ref{sec:threats}). One plausible explanation is that, as synthesis begins from empty (i.e.\ \texttt{true}) invariants, direct synthesis examples reflect the first step of the process, while mutation-based examples teach the model to repair failing invariants.

\begin{rqanswer}{3}
Supplementing mutation-based data with direct synthesis examples ($G_{D+M_1+M_2+M_4}$, 48.1\%/50.4\%) yields a small, non-significant improvement over mutation-based data alone ($G_{M_1+M_2+M_4}$, 45.8\%/47.9\%). However, the combination significantly outperforms direct synthesis alone ($G_D$, 44.3\%/45.6\%) under iterative refinement, suggesting that mutation increases the performance obtainable from a fixed set of verified programs, although training budgets differ (Section~\ref{sec:threats}).
\end{rqanswer}

\noindent\textbf{RQ4:} Increasing the LoRA rank and alpha from 8 to 32 ($G^{32}_{D+M_1+M_2+M_4}$) further improves performance to 50.8\%/53.6\%, a gain of 2.7/3.2 percentage points (2.2/3.5 at the method level), although neither gain is significant after correction. This is the best-performing Gemma configuration across both metrics and settings, verifying 6.3$\times$ as many programs as the base model zero-shot (3.2$\times$ under iterative refinement), and 10.6$\times$ as many eligible methods zero-shot.

\begin{rqanswer}{4}
Increasing the LoRA rank and alpha from 8 to 32 ($G^{32}_{D+M_1+M_2+M_4}$, 50.8\%/53.6\%) improves on the rank-8 configuration ($G_{D+M_1+M_2+M_4}$, 48.1\%/50.4\%) across both metrics and settings, though not significantly, suggesting that a rank of 8 may limit how much Gemma benefits from our training data.
\end{rqanswer}

\noindent\textbf{RQ5:} Fine-tuning on the combined data ($Q_{D+M_1+M_2+M_4}$) raises Qwen's performance from 9.3\%/12.3\% to 53.6\%/54.4\%, a 5.8$\times$ improvement zero-shot (4.4$\times$ under iterative refinement). Although base Qwen performs slightly better than base Gemma zero-shot (9.3\% compared to 8.1\%), it performs worse under iterative refinement (12.3\% compared to 16.5\%), gaining only 3.0 percentage points from iteration compared to Gemma's 8.4. Once fine-tuned, however, Qwen exceeds the equivalent Gemma configuration by 5.5/4.0 percentage points (3.1/2.8 at the method level), significantly so in the zero-shot setting ($p = 0.008$; $p = 0.056$ under iterative refinement), and also numerically exceeds $G^{32}_{D+M_1+M_2+M_4}$ at the program level despite using the default rank of 8. $Q_{D+M_1+M_2+M_4}$ achieves the highest program-level results of all configurations, while $G^{32}_{D+M_1+M_2+M_4}$ achieves the highest method-level result under iterative refinement.

\begin{rqanswer}{5}
Our approach generalises to a different model family: fine-tuning raises Qwen's performance from 9.3\%/12.3\% ($Q'$) to 53.6\%/54.4\% ($Q_{D+M_1+M_2+M_4}$), outperforming the equivalent Gemma configuration ($G_{D+M_1+M_2+M_4}$, 48.1\%/50.4\%) on both metrics, significantly so in the zero-shot setting.
\end{rqanswer}

\noindent\textbf{Iterative refinement.} Across all configurations, iterative refinement improves on zero-shot generation, though the size of the gain varies considerably. Base Gemma benefits substantially from iteration (8.4 percentage points for $G'$ and 16.1 for $G'_T$), whereas for fine-tuned models the gain ranges from 0.8 to 3.1 points. The smaller gains for fine-tuned models may partly reflect their stronger zero-shot performance, which leaves fewer instances for refinement to recover. However, base Qwen also gains comparatively little (3.0 points; RQ5) despite its weak zero-shot performance, suggesting that the ability to exploit verifier feedback also varies between model families. Moreover, as each training example captures a single refinement step, models are not explicitly trained to make progress across multiple iterations; training on multi-step refinement trajectories is a promising direction for future work.

\section{Threats to Validity}\label{sec:threats}
\noindent\textbf{Internal validity.} Each configuration was fine-tuned and evaluated once. Greedy decoding removes sampling variance at inference time, but results may still vary across training runs, so small differences between configurations, several of which are not statistically significant (Section~\ref{sec:results}), should be interpreted with caution. Our significance tests capture variation across test programs, but not across training runs. Hyperparameters were fixed across configurations rather than tuned for each, and the increased LoRA rank was evaluated for a single configuration only. Combining datasets increases both the volume and the diversity of training data, so RQ2 and RQ3 do not isolate either factor; the direct synthesis dataset is also smaller than each mutation-based dataset. We did not compare against direct synthesis examples drawn from additional source programs at matched volume; our results therefore show that mutation increases the data and performance obtainable from a fixed set of programs, rather than that mutants are more valuable per example than additional ground-truth invariants. Training budgets also differ: one epoch over the combined data involves roughly seven times as many optimisation steps over the same ground-truth invariants as one epoch over $D$, so part of the improvement of $G_{D+M_1+M_2+M_4}$ over $G_D$ may reflect longer training rather than the mutation-based prompts themselves. Finally, verification outcomes depend on the time limits used (Section~\ref{verification settings}): as timeouts are treated as failures, some sufficient invariants may have been rejected.

\noindent\textbf{Construct validity.} We measure the proportion of eligible methods and programs that verify. Successful verification establishes that the synthesised invariants are sufficient, assuming the soundness of Dafny and its underlying solver, but does not capture other qualities such as readability or similarity to human-written invariants. As the framework inserts only invariant clauses in the expected format, leaving the program and its specifications unchanged, verification cannot be achieved by weakening the specifications.

\noindent\textbf{External validity.} Our evaluation is limited to Dafny, a single benchmark, and two model families of similar size. The mutation operators are designed for Dafny invariants, and results may differ for other verifiers, languages, or model sizes. As DafnyBench has been publicly available since 2024, its programs may appear in the pretraining data of the models we evaluate. However, the low performance of the base models suggests that memorisation does not explain our results. Our fine-tuning data is drawn from a separate, synthetically generated dataset, and a comparison of training and test programs found no near-duplicates: excluding comments and loop invariants, the highest token-level similarity between any test program and its closest training program is 0.12 (Jaccard similarity over 5-token sequences).

\section{Related Work}
\label{sec:related}
Loop invariant synthesis is a central challenge in automated program verification~\cite{furia2014loop}. Since the problem is undecidable in general, no complete solution can exist; nonetheless, substantial prior work has approached it through a variety of methods.

Traditionally, techniques for loop invariant synthesis span dynamic and static program analyses. Dynamic approaches infer loop invariants from program execution traces. For example, Daikon~\cite{ernst2007daikon} attempts to derive likely invariants for a given program from its test traces. DIDUCE~\cite{hangal2002tracking} and DIG~\cite{nguyen2012using} also infer likely invariants from observed executions, the latter including nonlinear polynomial and array relations. In contrast, static approaches attempt to reason about loop invariants without program execution. A wide array of techniques in this category has found application to invariant synthesis: abstract interpretation~\cite{cousot1977abstract}, logical abduction~\cite{calcagno2009bi}, recurrence analysis~\cite{humenberger2017invariant}, and symbolic execution~\cite{nguyen2017symlnfer,tabar2022automatic}. Furia and Meyer~\cite{furia2010inferring} derive candidate loop invariants from a loop's postcondition. Some approaches combine both: DynaMate~\cite{galeotti2015inferring} mutates postconditions to obtain candidate invariants, filters them using dynamic analysis, and verifies the remainder with a static checker. Mutation testing has also been combined with genetic programming to search for invariants~\cite{ratcliff2011searching}.

Beyond program analyses, learning-based approaches utilise machine learning in some capacity to synthesise loop invariants. Many of them are data-driven: they learn candidates from program states and check them with a verifier, whose counterexamples become new data. Guess-and-Check~\cite{sharma2013data} learns algebraic equality invariants in this way, ICE~\cite{garg2014ice} also learns from implication counterexamples, PIE~\cite{padhi2016data} learns the features of candidate invariants on demand, and NumInv~\cite{nguyen2017counterexample} checks numerical invariants with the KLEE test generator. Garg et al.~\cite{garg2016learning} extend ICE with a decision tree learning algorithm; Code2Inv~\cite{si2020code2inv} relies on deep reinforcement learning (RL) and proof-checker interactions; and LIPuS~\cite{yu2023loop} utilises RL-based pruning and SMT solving. Haltermann and Wehrheim~\cite{haltermann2022machine} re-implement four ML-based invariant generators in a common framework to compare them and study their reproducibility.

More recently, researchers have applied language models to specification, annotation, and loop invariant synthesis~\cite{king2025llm, king2026llms, endres2024can, ma2025specgen, kamath2023finding, pei2023can, wu2023lemur, granberry2025specify, beg2026evaluating}. Some approaches rely on zero-shot generation~\cite{janssen2024can}, whilst others follow a \textit{generate-and-check} paradigm~\cite{wu2024llm}. For example, Pirzada et al.~\cite{pirzada2024llm} use an LLM to generate invariant candidates, which are then validated by the Vampire theorem prover~\cite{kovacs2013first} and used to replace loops for bounded model checking~\cite{clarke2001bounded}. Language models have also been trained on failed proofs. Baldur~\cite{first2023baldur} fine-tunes an LLM to generate whole proofs for a proof assistant, and fine-tunes a second model to repair a failed proof given the failed attempt and its error message. An et al.~\cite{an2024learn} train a model on failed search paths as well as successful ones, and find that it proves more unseen theorems than a model trained on successful paths alone.

Of particular interest are works which concern Dafny. Faria et al.~\cite{pascoal2025automatic} use repeated prompt-based sampling with verifier-based validation for invariant synthesis. AxDafny~\cite{breen2026axdafny} presents an agentic, verifier-guided approach that uses frontier LLMs to generate implementations and proof annotations for given specifications. Laurel~\cite{mugnier2025laurel} concerns generating assertions, and presents a proof-similarity metric for retrieval-augmented generation. Silva et al.~\cite{silva2024leveraging} propose using LLMs to help Dafny developers, for example by suggesting missing lemmas, and DAISY~\cite{silva2025inferring} uses LLMs to infer missing helper assertions, locating them by combining LLM predictions with error-message heuristics. Wu et al.~\cite{wu2026specification} use Dafny specifications as oracles to localise arithmetic bugs, and repair them with LLMs.

Furthermore, fine-tuning language models for invariant synthesis is a growing research area. Pei et al.~\cite{pei2023can} fine-tune language models of code for invariant generation, and find that they can predict invariants statically, without execution traces. DafnyPro~\cite{banerjee2026dafnypro} presents an inference-time framework for generating proof annotations (including loop invariants) in Dafny, and fine-tunes Qwen2.5-7B and Qwen3-14B for this task. Its training data pairs failed attempts from its framework with the verified ground-truth annotations and informal explanations of the verifier errors. The dafny-annotator tool~\cite{poesia2024dafny} fine-tunes LLaMA 3.1 8B and CodeLlama 7B to add annotations to Dafny programs, training on part of DafnyBench and testing on held-out DafnyBench programs. ATLAS~\cite{baksys2025atlas} synthesises verified Dafny programs from programming problems with Python solutions, and fine-tunes Qwen2.5-Coder-7B on tasks extracted from them, including the repair of failed attempts. Pinto et al.~\cite{pinto2026not} present Wonda, a data curation pipeline used to train SLMs across the Qwen3, Llama-3.1, and Mistral model families for invariant synthesis in C. Likewise, Quokka~\cite{wei2025quokka} presents both an LLM-based approach to invariant synthesis in C and a data generation method for fine-tuning. Additionally, Re:Form~\cite{yan2025re} and Formal Disco~\cite{poesia2026formal} both present methods for generating synthetic datasets of Dafny programs. Re:Form trains models of 0.5B to 14B parameters to generate Dafny specifications, using reinforcement learning with a reward from the Dafny verifier, while Formal Disco fine-tunes Qwen2.5-Coder-32B to generate proof annotations.

Our work focuses on the training data used to fine-tune SLMs for invariant synthesis. Wonda~\cite{pinto2026not} curates correct invariants as training targets, while Re:Form~\cite{yan2025re} and Formal Disco~\cite{poesia2026formal} generate new programs. DafnyPro~\cite{banerjee2026dafnypro}, ATLAS~\cite{baksys2025atlas}, and Baldur~\cite{first2023baldur} also train on failed attempts paired with error feedback, but they collect these failures from runs of language models. We instead produce failing invariants systematically through semantic mutation, augmenting an existing dataset of verified Dafny programs. Each failing invariant is paired with verifier feedback and its ground-truth counterpart, training models for feedback-guided invariant refinement. Training on synthetically corrupted inputs is established in program repair, where DeepFix~\cite{gupta2017deepfix}, Break-It-Fix-It~\cite{yasunaga2021break}, and BugLab~\cite{allamanis2021self} learn to fix errors introduced into correct code; we apply this idea to loop invariants, using the verifier both to confirm that each corruption induces failure and to provide feedback. While mutation has previously been used in invariant inference~\cite{galeotti2015inferring, ratcliff2011searching}, specification generation~\cite{ma2025specgen}, the testing of Alloy specifications~\cite{sullivan2017automated}, and the creation of training data for neural bug detectors~\cite{richter2022learning}, to our knowledge ours is the first approach to use it to generate training data for invariant synthesis. Galeotti et al.\ identify the blow-up in the number of mutants as a fundamental challenge for inference~\cite{galeotti2015inferring}; when generating training data, this abundance instead becomes an asset. Finally, unlike approaches that rely on frontier LLMs at inference time~\cite{pascoal2025automatic, breen2026axdafny}, our fine-tuned SLMs are small enough to run locally.

% Similarly, the work of~\cite{wu2024llm} uses both LLMs and bounded model checking: their method employs LLMs to generate loop invariant candidates and bounded model checking to filter incorrect predicates. The Lemur framework~\cite{wu2023lemur} uses LLMs to generate and repair candidate loop invariants, which are then used within the ESBMC~\cite{menezes2024esbmc} and UAutomizer~\cite{heizmann2016ultimate} frameworks for program verification.

\section{Conclusion and Future Work}\label{sec:conclusion}
We presented a failure-driven data curation method for fine-tuning SLMs to synthesise loop invariants in Dafny. Inspired by mutation testing, our method applies semantic mutation operators to ground-truth invariants, producing verification-failing invariants that, paired with verifier feedback and their ground-truth counterparts, serve as training examples for an invariant refinement framework. Whereas the blow-up in the number of mutants is an obstacle in mutation testing, it becomes an asset when generating training data, multiplying the examples obtainable from a fixed set of verified programs. Our best-performing Gemma-4-E4B configuration raised the base model's zero-shot verification rate on DafnyBench from 8.1\% to 50.8\%, with higher-order mutants proving more effective than first-order mutants, and the addition of mutation-based examples improving on direct synthesis examples alone under iterative refinement. The approach also transferred to Qwen3.5-4B, raising its zero-shot verification rate from 9.3\% to 53.6\%. These results show that systematically generated training data can make SLMs, which run locally and reproducibly, substantially more capable invariant synthesisers.

Several directions remain for future work. As each training example captures a single refinement step, training on multi-step refinement trajectories may help models make better use of iteration. Ablations isolating the effects of data volume and diversity, and of individual mutation operators, would clarify which failures are most instructive. Mutation-based examples could also be combined with failures collected from model outputs, as in DafnyPro~\cite{banerjee2026dafnypro} and ATLAS~\cite{baksys2025atlas}, to cover both systematic and model-specific errors. Finally, as our approach requires only ground-truth annotations and a verifier, it could be extended to other annotation types, such as assertions and termination measures.

\section*{Data Availability}
Our supplementary material~\cite{supp} includes all training datasets, the test dataset, logs of the models' interactions during evaluation, and an example refinement prompt with its training target.

\section*{Acknowledgement}
The authors used generative AI (Claude Opus 5.5) to assist with identifying grammatical errors and inconsistencies; improving prose; checking that cited works are referenced accurately; writing automation scripts for running experiments and collating results; and producing figures and tables. All technical content, research decisions, and interpretations are the authors' own. All AI-assisted text and code was reviewed and edited by the authors.

\bibliographystyle{IEEEtranDOI} % VK this will show DOIs
\bibliography{references.bib}

\end{document}